# HyMNet: a Multimodal Deep Learning System for Hypertension Classification using Fundus Photographs and Cardiometabolic Risk Factors

**Mohammed Baharoon[a,b], Hessa Almatar[a,b], Reema Alduhayan[a,b], Tariq Aldebasi[c], Badr Alahmadi[d], Yahya Bokhari[a,b], Mohammed Alawad[a,b,e], Ahmed Almazroa[a,b,†], Abdulrhman Aljouie[a,b,†]**

[a] AI and Bioinformatics Department, King Abdullah International Medical Research Center, Riyadh, Saudi Arabia

[b] King Saud bin Abdulaziz University for Health Sciences (KSAU-HS), Riyadh, Saudi Arabia

[c] Ophthalmology Department, King Abdulaziz Medical City, Ministry of National Guard Health Affairs, Riyadh, Saudi Arabia

[d] Ophthalmology Department, Prince Mohammad bin Abdulaziz Hospital, Ministry of National Guard Health Affairs, Al Madinah, Saudi Arabia

[e] National Center for Artificial Intelligence (NCAI), Saudi Data and Artificial Intelligence Authority (SDAIA), Riyadh, Saudi Arabia

**Correspondences:** Ahmed Almazroa and Abdulrhman Aljouie
**Email addresses:** almazroaah@ngha.med.sa, aljouieab@ngha.med.sa
**† Corresponding Authors**

## Abstract

In recent years, deep learning has shown promise in predicting hypertension (HTN) from fundus images. However, most prior research has primarily focused on analyzing a single type of data, which may not capture the full complexity of HTN risk. To address this limitation, this study introduces a multimodal deep learning (MMDL) system, dubbed *HyMNet*, which combines fundus images and cardiometabolic risk factors, specifically age and gender, to improve hypertension detection capabilities. Our MMDL system uses RETFound, a foundation model pre-trained on 1.6 million retinal images, for the fundus path and a fully connected neural network for the age and gender path. The two paths are jointly trained by concatenating the feature vectors from each path that are then fed into a fusion network. The system was trained on 5,016 retinal images from 1,243 individuals collected from the Saudi Ministry of National Guard Health Affairs. The results show that the multimodal model that integrates fundus images along with age and gender outperforms the unimodal system trained solely on fundus photographs, with an F1 score of 0.771 [0.747, 0.796], and 0.745 [0.719, 0.772] for hypertension detection, respectively. Additionally, we studied the effect underlying diabetes mellitus has on the model's predictive ability, concluding that diabetes is used as a confounding variable for distinguishing hypertensive cases. Our code and model weights are publicly available at https://github.com/MohammedSB/HyMNet.

**Abbreviations**

BP, blood pressure; CVD, cardiovascular disease; EHR, electronic health record; EMR, electronic medical records; AI, artificial intelligence; DL, deep learning; MMDL, multimodal deep learning; SVM, support vector machine; FCNN, fully connected neural network; CNN convolutional neural network; ReLU; rectified linear unit; AUC, area under the operating characteristic curve, PR, area under the precision-recall curve; CI, confidence interval; MAE, mean absolute error; KAIMRC, King Abdullah International Medical Research Center.

**1. Introduction**

Cardiovascular diseases persist as one of the primary causes of mortality worldwide, with hypertension, or high blood pressure (BP), serving as a significant contributing risk factor (1,2). High BP is responsible for approximately 54% of stroke occurrences and 47% of coronary heart disease incidents (3). Moreover, high BP increases the likelihood of developing hypertension-mediated organ damage, such as retinopathy and renal failure (4,5). Despite the severe and life-threatening consequences of hypertension, nearly 46% of adults with high BP remain unaware of their condition (6). Consequently, there is a pressing need for tools that facilitate early detection and identification of hypertension, which can aid in CVD risk stratification and prevent further complications arising from the condition (7).

Outpatient blood pressure measurements may not accurately represent a patient's true blood pressure, leading to potential over- or underestimation and subsequently impacting proper cardiovascular disease risk stratification (8–11). A single high blood pressure reading might merely be a manifestation of stress, known as white coat syndrome, and not indicative of an individual's true chronic blood pressure state (12–14). Conversely, a normal blood pressure measurement could provide false reassurance, masking underlying hypertension (14–16). Since hypertension can affect microvascular structures early in the disease process, long before clinical signs and symptoms become evident, assessing microvascular health may offer a more accurate representation of CVD risk in an outpatient setting (17,18).

High blood pressure can contribute to microvascular damage in its earlier stage, causing alterations to small blood vessels, such as narrowing and rupturing (19). Specifically, within the retina, high BP can initially result in focal arteriolar narrowing and arteriovenous nipping, progressing to lipid exudation (visible as hard exudates) and ischemia of nerve-fiber layers (cotton-wool spots) in later or more severe stages (17). The retina is unique in that it allows for non-invasive visualization of vasculature (20,21). Consequently, retinal fundus photography can capture the vascular changes induced by hypertension, positioning it as a promising method for early hypertension identification and screening (21–23).

Deep learning encompasses a group of machine learning algorithms designed to automatically learn and extract complex patterns and features from vast amounts of data (24). These algorithms can capture subtle details and relationships within fundus images that may not be immediately apparent to the human eye, thereby enhancing their ability to detect microvascular changes. By effectively processing and analyzing these intricate visual cues, DL models can potentially outperform traditional diagnostic methods and offer more accurate and efficient identification of hypertension, specifically in its early stages. In the past, DL has demonstrated exceptional performance in classifying retinal diseases, such as diabetic retinopathy, hypertensive retinopathy, and glaucoma, using fundus photographs (25–28).

Multimodal deep learning involves the use of heterogeneous data modalities to train deep learning systems. Similar to physicians who base decisions on inputs from various sources (e.g., physical examination, patient history, lab results), DL systems must also incorporate data from multiple modalities to achieve clinician-level accuracy. Recent studies have shown that MMDL can improve predictive performance for CVD risk assessment and CVD detection by utilizing these diverse data inputs (29–31).

## 1.1. Background

Prior research studies have explored the application of DL for hypertension classification by utilizing fundus photographs. In their study, Zhang and colleagues utilized an ImageNet pre-trained Inception-v3 neural network, which is a type of Convolutional Neural Network (CNN), to classify hypertension based on fundus photographs. Specifically, they focused on a two-class classification task, distinguishing between individuals with hypertension (defined as SBP > 140 or DBP > 90 mmHg) and those with normal blood pressure. They evaluated the performance of their approach using a dataset of 1,222 fundus images from a population in Central China, and reported an area under the receiver operating characteristic curve of 0.766 for hypertension classification (32). Poplin et al. demonstrated the potential of deep learning in extracting valuable cardiometabolic risk factors from retinal images (33). Their model, based on the Inception-v3 architecture, was trained on a large dataset of 1,779,020 images from EyePACS and the UK Biobank. The model was able to accurately extract age with a mean absolute error of 3.26 years, gender with an area under the curve (AUC) of 0.97, smoking status with an AUC of 0.71, systolic and diastolic blood pressures with MAE values of ±11.35 and ±6.42 mmHg respectively, and body mass index with an MAE of ±3.29. Building upon the research of Poplin et al., Gerrits et al. conducted a study to investigate the potential mediating effects of age and gender on the predictive performance of MobileNet-V2, a deep learning architecture, in assessing cardiometabolic risk factors such as hypertension, using a dataset of 12,000 fundus photographs from the Qatar Biobank (34). Their findings indicate that age and gender may act as mediating variables when predicting blood pressure and other cardiometabolic risk factors.

The present study expands upon the work of Zhang et al. and Gerrits et al (32,34) by proposing a multimodal DL system that integrates fundus photographs with cardiometabolic risk factors using various data fusion techniques. Age and gender were deliberately chosen as supplementary features to fundus photographs, given their ease of accessibility and role as risk factors for hypertension. As a result, the proposed MMDL system has the potential to assess hypertension risk in ophthalmology clinics and serve as a screening tool for early hypertension identification.

## 2. Methods

## 2.1. Dataset and Label Distribution

The data used in this study was acquired from KAIMRC's big ocular images dataset (35). The collection of these images was approved by the Ministry of National Guard Health Affairs (MNGHA) institutional review board (IRB) under protocol number RC-19-316-R. Since the images were fully anonymized and collected retrospectively, the IRB waived the need for obtaining patients' informed consent. The study is conducted in adherence to the principles outlined by the Declaration of Helsinki.

We selected 5,016 fundus images from 1,243 patients. All the patients' characteristics were summarized in Table 1. Each patient had a range of fundus images, with a minimum of one image and a maximum of 43 images. A total of 1,035 patients had images for both the right and left eyes, while 195 patients have an image for only one eye. The fundus images were collected retrospectively from patients using two versions of TOPCON OCT machines (DRI OCT Triton and 3D OCT-2000) and originally had dimensions of 2576x1934 pixels.

Each observation in the dataset is defined by its demographic attributes, including age and gender, as well as fundus photographs and a binary classification of hypertension status: either hypertensive or non-hypertensive. The classification was based on the patient's history of having hypertension, using antihypertensive treatments, or at least three readings of high blood pressure. The information was extracted from the patient's Electronic Medical Record (EMR) progress notes written by the ophthalmologist on the same visit date for fundus image capturing.

The dataset was separated into training, validation, and test sets. Roughly 60% (n=745) of the dataset was dedicated to model training, 20% (n=249) was allocated for model selection and hyperparameter tuning, and the remaining 20% (n=249) was reserved for testing the model's performance. We divided the data so that the same ratio of hypertensive to non-hypertension patients exists across the three subsets. Additionally, we made sure that the ratio of hypertensive to diabetic patients is consistent across all three subsets. To ensure that no patient appeared in multiple subsets, a patient-specific split was carried out in which patients with both right and left eye images were allocated to either the training, validation, or test set. This approach ensured that each patient was exclusively represented in a single subset.

### 2.1.1. Descriptive Analysis

A descriptive analysis of the dataset shows that the average age of the subjects was 59.4±29.0. Hypertensive individuals had an average age of 63.4±10.2, while non-hypertensive individuals had an average age of 52.7±14.99. The dataset was composed of 44% male (n=544) and 56% female (n=699) subjects. 703 were hypertensive and the remaining 540 were non-hypertensive. Moreover, of the 1,243 patients, 978 patients had diabetes mellitus. Focusing just on the 703 patients with hypertension, 668 had diabetes (95%). Further details on the patient characteristics of each gender can be found in Table 1.

Table 1. Patient Characteristics. The mean and standard deviation for hypertensive, non-hypertensive, and both categories combined are presented for each gender. The percentage of male and female patients in the entire dataset is also shown. We report percentages for diabetic patients divided by gender and hypertension status.

| Patient Characteristics | Age (mean±std) [CI 95%] | | Diabetes | | Gender Split | |
|---|---|---|---|---|---|---|
| | Male | Female | Male | Female | Male | Female |
| Hypertension | 63.7±11.5 [62.4, 65.0] | 62.8±9.1 [62.0, 63.8] | 95% (n=295) | 95% (n=373) | 44% (n=310) | 56% (n=393) |
| Non-hypertension | 54.2±15.0 [52.3, 56.2] | 51.5±14.8 [49.9, 53.2] | 55% (n=130) | 58% (n=180) | 44% (n=234) | 56% (n=306) |
| All | 59.6±13.9 [58.4, 60.8] | 57.9±13.2 [56.9, 58.9] | 78% (n=425) | 79% (n=553) | 47% (n=544) | 53% (n=699) |

### 2.1.2. Data Preprocessing

Images were cropped and resized to 512x512 pixels, normalized to be in the range [0, 1] using min-max normalization, and standardized using the z-score formula with the mean and standard deviation of ImageNet. The age feature was also standardized to having a mean of zero and unit standard deviation.

### 2.1.3. Image Augmentation

We used standard image augmentation techniques to prevent model overfitting on the limited training data. Specifically, we performed rotation, flipping, and blurring, which are commonly used for training deep learning models.

During the training process, patient images were randomly rotated up to 360 degrees and flipped horizontally during the training process. Additionally, a random Gaussian blur with a kernel size of 3 was applied to further introduce small variations in the data set.

## 2.2. Classification Models

Four multimodal deep learning (DL) systems were devised to classify hypertension by integrating fundus photographs and demographic features with intermediate and late fusion techniques. Detailed descriptions of each system can be found in Section 2.2.1. The systems are comprised of three primary neural network components: a "*FundusPath*" that processes fundus photographs, a "*DemographicPath*" that handles age and gender features, and a "*FusionPath*" that integrates features from both modalities. A diagram for the four systems is represented in Appendix A, Supplementary Figure 1.

For the *FundusPath*, we employed RETFound, a foundation model pre-trained on 1.6 million retinal images from various sources (36). For the *DemographicPath*, we utilized a fully connected neural network (FCNN) consisting of two layers when using joint fusion, and four layers when using late fusion. Leaky Rectified linear activation functions (Leaky ReLU) (37) were used, and dropout layers were added (38). Just like the *DemographicPath*, the *FusionPath* is also a FCNN made up of four layers, that receives outputs from both paths and combines them to generate a hypertension prediction.

We additionally evaluated four unimodal systems for processing either fundus photographs or demographic features, detailed in Section 2.2.2.

### 2.2.1. Multimodal Systems

Here we explain the four multimodal deep learning systems *IntermediateFusion, PredictionFusion*, *LateFusion*, and *VotingFusion*. We provide a visual diagram for each method in Supplementary Figure 2.

**IntermediateFusion and PredictionFusion**

For the *IntermediateFusion* system, shown in Figure 1, the *FundusPath* extracts 8 features from the fundus photographs, while the *DemographicPath* outputs 32 deep features from the demographic information. The feature vectors generated by both networks are then concatenated, forming a composite feature representation. This composite representation was subsequently fed into the *FusionPath* network to produce hypertension classification. We choose the number of features outputted from each path by empirically testing different values, showing an ablation in Supplementary Figure 1.

For the *PredictionFusion* system, fundus images along with age and gender were passed into the *FundusPath* and *DemographicPath,* respectively. Unlike feature fusion, each path now produces a single output, which we call the prediction logit. Consequently, the prediction logits—rather than the deep features—from both paths were concatenated and passed into the *FusionPath*.

In both systems, the three networks that constitute the *FundusPath*, *DemographicPath*, and *FusionPath* were trained jointly. We will refer to the *IntermediateFusion* system as *HyMNet*.

**LateFusion**

Here we combined the prediction logit obtained from a fully trained RETFound, which was trained to predict hypertension using fundus photographs, with age and gender features. This concatenated data was then fed into a classifier. We evaluated three different classifiers: an XGBoost, a support vector machine (SVM), and a fully connected neural network (FCNN).

**VotingFusion**

For the *VotingFusion* method, we derived a prediction logit from fundus photographs using a trained RETFound model, as well as a prediction logit from demographic features utilizing a trained FCNN. Additionally, we acquired a third prediction logit from a fully trained *IntermediateFusion* system. These three prediction logits were then concatenated and fed into three separate classifiers: an XGBoost, an SVM, and an FCNN. Furthermore, we assessed an ensemble technique on the three prediction logits through soft voting, where the ultimate prediction would be the average of the three individual predictions.

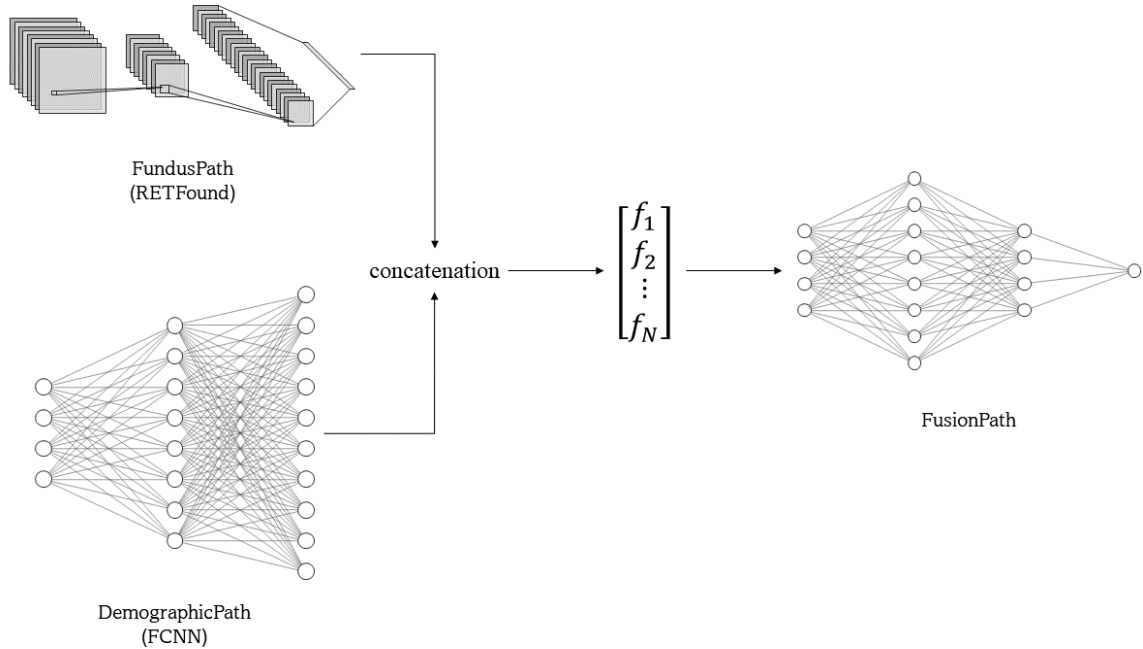

**Figure 1.** IntermediateFusion system diagram. Deep feature outputs from the FundusPath and the DemographicPath, represented as $f_1 - f_N$, are concatenated and fed into the FusionPath. The FusionPath's output is used to update all the trainable parameters for all three networks.

### 2.2.2. Unimodal Systems

The study aimed to compare the effectiveness of multimodal and unimodal hypertension models. To this end, we evaluated the performance of the RETFound, a unimodal model using fundus photographs. Additionally, we constructed a model using demographic features (age and gender) with three classifiers – XGBoost "*DemographicXGB*", a support vector machine "DemographicSVM", and a fully connected neural network "*DemographicFCNN*".

### 2.3. Training Configurations

We used the PyTorch framework for our deep learning pipeline (39). SVM modeling, data handling procedures, and vectorized operations were facilitated through the use of the scikit-learn, Pandas, and NumPy libraries (40–42). The XGBoost library was utilized for implementing the XGBoost classifier (43). During neural network training, we used a binary cross-entropy loss function and an AdamW optimizer (44). All training processes were conducted with a batch size of 16. We used 25 training epochs for validation runs and 50 epochs for testing all systems, except in only FCNN models, where we used 250 epochs. We selected the optimal checkpoint across all epochs. Furthermore, we employed a cosine scheduler to decrease the learning rate throughout each iteration.

## 2.4. Model Selection and Hyperparameter Tuning.

To determine the optimal neural network architectures and hyperparameters for the MMDL systems' main components, we conducted several experiments on the validation set using the AUC score as our performance metric.

To select the best architecture for the *FundusPath*, we tested various CNN and ViT architectures, including ResNet50, DenseNet-201 pre-trained on ImageNet1k, DINOv2 pre-trained on LVD-142M, and RETFound pre-trained on 1.6 million retinal images, with different learning rates (45–50). From these experiments we selected the best performing model for our *FundusPath*. The results for each network are presented in Section A.1 in the supplementary material.

For the *DemographicPath* and *FusionPath*, we experimented with different numbers of layers and learning rates. We also used cross-validation techniques to perform hyperparameter tuning for selecting the most optimal XGBoost and SVM parameters for each system.

In Section A.2 in the supplementary material, we studied the effect of fundus image size and concluded that increasing the size from 224x224 to 512x512 results in slightly better performance. Moreover, with the *IntermediateFusion* system, we further examined the effect of the feature vector size from both the *FundusPath* and the *DemographicPath* and presented our finding in Section A in the supplementary material. A decision threshold of 0.5 was used for generating the classifications of hypertension performance metrics.

## 2.5. Statistical Analysis

We employed a non-parametric bootstrapping method to assess the statistical significance of our findings. This approach involved resampling the test set with replacement, where the number of samples taken was equal to the total number of observations in the test set. By evaluating the model's performance on these resampled datasets, we gained insights into its true capabilities.

To obtain a robust estimate of the model's performance, we repeated this resampling procedure 10,000 times, which is used to balance computational efficiency and accurate parameter estimation. For each iteration, we calculated the performance metrics, such as the AUC, PR, and other metrics. To determine the 95% confidence intervals for these metrics, we reported the values at the 2.5th and 97.5th percentiles. This approach provided a range within which we could expect the model's true performance to fall, accounting for the variability in the data and the potential impact of sampling bias.

## 2.6. Experimental Environment

We conducted all the computational experiments with a workstation that has an AMD Ryzen Threadripper PRO 5955WX 16-Cores processor and an NVIDIA RTX A6000 GPU.

## 3. Results

The results of our study indicate that incorporating retinal images and age and gender features improves hypertension prediction capability.

First, we focus our analysis on deep learning models, excluding SVM and XGBoost models. In Table 2, we compare RETFound and the *DemographicFCNN* model to *HyMNet*. Notably our *HyMNet* approach demonstrated the highest scores across all five metrics, with an F1 score of 0.771 [0.747, 0.796] for *HyMNet*, while an F1 score of 0.745 [0.719, 0.772] and 0.752 [0.727, 0.778] were achieved by RETFound and *DemographicFCNN*, respectively. Figure 2 presents the F1 score of the three systems on a box and whisker plot. These results underscore the significance of including demographic features in conjunction with fundus photographs to improve the performance of hypertension prediction models.

**Table 2. Multimodal and unimodal systems comparison.** The table presents the F1-score, AUC, PR, precision, recall, and specificity scores for the HyMNet, RETFound, and DemographicFCNN with CI representing the 95% confidence interval generated from the bootstrap technique mentioned in Section 2.5. A classification threshold of 0.5 was used for the F1-score, precision, recall, and specificity.

| Model | F1 | AUC | PR | Accuracy | Precision | Recall |
|---|---|---|---|---|---|---|
| HyMNet | **0.771 [0.747, 0.796]** | **0.705 [0.672, 0.738]** | **0.743 [0.703, 0.784]** | **0.690 [0.662, 0.719]** | **0.683 [0.65, 0.716]** | **0.887 [0.862, 0.912]** |
| RETFound | 0.745 [0.719, 0.772] | 0.690 [0.657, 0.724] | 0.740 [0.701, 0.78] | 0.682 [0.647, 0.717] | 0.668 [0.639, 0.698] | 0.821 [0.791, 0.852] |
| DemographicFCNN | 0.752 [0.727, 0.778] | 0.694 [0.661, 0.727] | 0.742 [0.703, 0.782] | 0.661 [0.632, 0.69] | 0.662 [0.63, 0.695] | 0.871 [0.845, 0.898] |

To evaluate the statistical significance of the F1 performance for *HyMNet* compared to the other systems, we employ the method used in (51). Specifically, we compute the difference between the bootstrapped F1 scores of *HyMNet* and RETFound using the identical bootstrap sample. Following this, we ascertain if the 95% CI for this difference encompasses zero or a negative value. We discover that the difference in F1 scores — 0.02 [0.003, 0.038] — excludes zero, leading us to conclude that *HyMNet*'s performance is statistically significantly better than that of RETFound. We repeat this experiment for the DemographicFCNN and conclude that the performance increase is not statistically significant, with a difference in F1 scores of 0.013 [-0.005, 0.032].

We further present additional results in Supplementary Table 2 for all the systems used in the study. We observe that combining *HyMNet* with RETFound and *DemographicFCNN* in the *VotingFusionEnsemble* method further increases the performance slightly across all metrics except Recall.

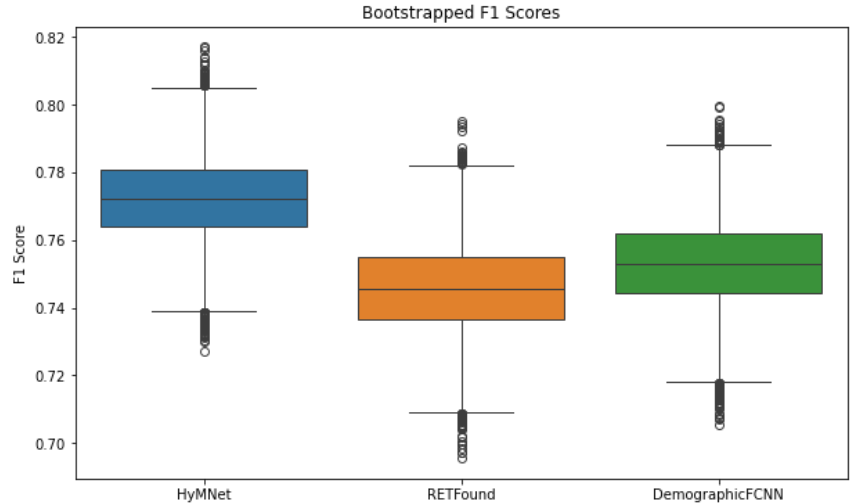

**Figure 2. Box and whisker plot for F1 score of HyMNet, RETFound, and DemographicFCNN.** This plot shows the increase in performance achieved by incorporating fundus photograph features with demographic features. The plot was generated using the 10,000 bootstrapped F1 scores.

In figure 3, we present the ROC and PR curves for *HyMNet*, RETFound, and *DemographicFCNN*. Specifically, we plot the bootstrap run with the median AUC and PR value (presented in the dark colors) and use the 97.5 and 2.5 percentile from the 10,000 bootstrapped results to plot the interval (highlighted in the lighter colors). We observe a large variability in both plots.

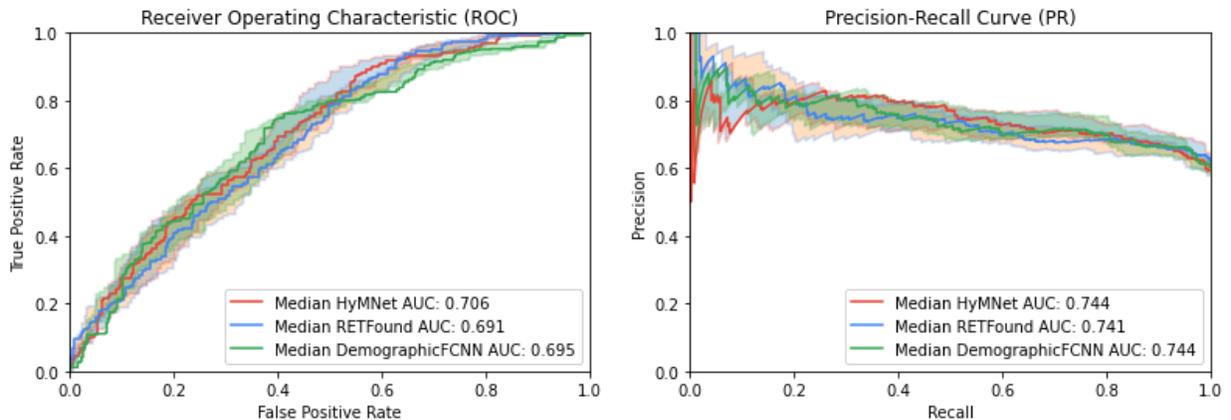

**Figure 3. ROC and PR curve comparison of multimodal and unimodal systems.** The figure shows ROC and PR curves for HyMNet, RETFound and DemographicFCNN. The diagrams are generated using the predictions of median area under the curve scores out of the 10,000 bootstrap runs. The 97.5 and the 2.5 percentiles ROC and PR curves are also shown using lighter colors.

### 3.1 The Influence of Diabetes on Hypertension Prediction

We also examine the impact of diabetes on the detection of hypertension. In Table 3, we display the performance measures for *HyMNet* when applied to patients with diabetes versus those without diabetes. We note that the F1 and PR scores are nearly 0.5 greater in patients with diabetes, suggesting that diabetes is a confounding variable for hypertension prediction. There were 849 (84%) and 158 (16%) images from diabetic and non-diabetic patients, respectively.

**Table 3. Presence of diabetes on hypertension detection capabilities.** The table presents performance metrics for HyMNet on patients with diabetes compared to patients without diabetes. CI represents the 95% confidence interval generated from the bootstrap technique mentioned in Section 2.5. A classification threshold of 0.5 was used for the F1-score, precision, recall, and specificity.

| Diabetes Status | F1 | AUC | PR | Accuracy | Precision | Recall |
|---|---|---|---|---|---|---|
| Positive | 0.796 [0.772, 0.821] | 0.68 [0.642, 0.717] | 0.788 [0.748, 0.828] | 0.696 [0.665, 0.727] | 0.716 [0.684, 0.749] | 0.895 [0.869, 0.921] |
| Negative | 0.466 [0.352, 0.581] | 0.704 [0.617, 0.79] | 0.306 [0.202, 0.411] | 0.642 [0.57, 0.715] | 0.344 [0.237, 0.45-1] | 0.78 [0.636, 0.923] |

## 3.2 Region of Interest Visualization

We employ Grad-CAM (52) to visualize the regions of interest used by RETFound to make predictions. We selected five hypertensive images at random and predicted their activations maps with a RETFound model that is further trained to predict hypertension. Figure 4 shows the resulted activation map, with red representing higher activations. The map shows that blood vessels and the disc margin are used by the model to make predictions. However, the regions of interests are not conclusive, and we did not clearly observe the expected pattern of primarily focusing on blood vessels (53).

**Figure 4. Grad-CAM for fundus photographs with hypertension.** We used RETFound for this analysis. The fundus images are selected at random.

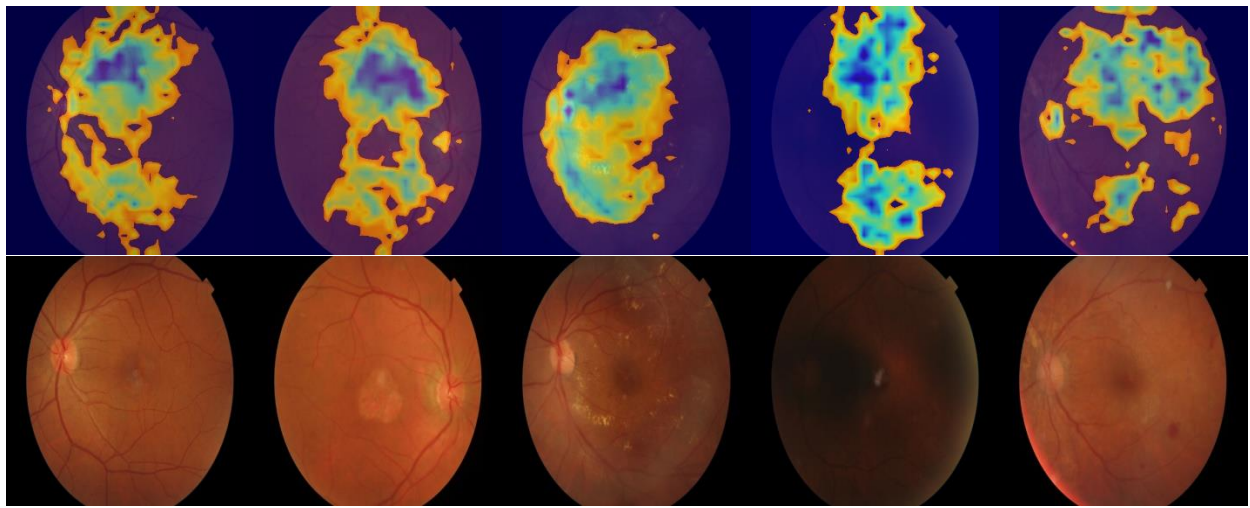

## 4. Discussion

Hypertension is a condition that can have various effects on the vascular system, which can be visualized through fundus photographs. However, these effects may occur at the microvascular level, making it difficult for human observers to detect them, especially in the early stages. The use of deep learning techniques may offer a solution to this challenge, as they exhibit high sensitivity to such changes, potentially enabling early detection of hypertensive patients and preventing the onset of severe health conditions. Previous studies have shown that age and gender are strong predictors of hypertension (54), with males being at a higher risk of developing hypertension at an earlier age compared to females (55). Given these findings, it is natural to explore the integration of readily available and robust sociodemographic predictors, such as age and gender, into the development of a multimodal model for hypertension that incorporates fundus photographs. This approach could lead to an effective and efficient model for early detection and management of hypertension, ultimately improving patient outcomes.

In the medical AI field, multimodal deep learning has recently emerged as a promising area of research (56). By utilizing diverse heterogeneous data, MMDL systems attempt to mimic medical expert's decision-making processes, which are typically based on a variety of sources. In a study by Qiu et al., deep learning models showed improved predictive performance for dementia and Alzheimer's disease identification when integrating magnetic resonance imaging (MRI) scans with demographic features, and other electronic health records (EHR) clinical data (57). Additionally, Lee et al. have demonstrated increased predictive performance for CVD risk using multimodal deep learning, by combining fundus photographs with cardiometabolic risk factors (58).

We conducted several experiments using *HyMNet*, which integrates age and gender information with fundus images. *HyMNet* achieved an F1 score of 0.77 in predicting hypertension, surpassing the F1 scores of the RETFound and *DemographicFCNN* models, which scored 0.74 and 0.75, respectively. This improved performance can be attributed to the fact that hypertension typically develops later in life and may have different onset patterns between males and females (55).

Our unimodal fundus model for hypertension prediction resulted in an AUC of 0.69, which is within range of values achieved by previous studies (32,54). Zhang et al. have investigated deep learning models for hypertension classification using fundus photographs where their model achieved an AUC of 0.77, utilizing 1,222 images from a central China population (32). However, Dai et al. achieved a lower AUC of 0.65 using a dataset of 2,012 images from an east Asian population, which could be attributed to factors such as data quality (59).

**Limitation and Future Direction**

The primary limitation of this study is the absence of hypertension stages in the dataset. Namely, we were not able to assess the capability of our model to predict hypertension at different stages, given that we don't have information about the hypertension stage for each patient. On the same note, we were not able to measure the effect hypertensive retinopathy has on the ability to predict systemic hypertension, and we believe that this is a promising direction for future research.

This study is also limited by a small dataset, which led to instability in the bootstrap performance. This is demonstrated by a 95% confidence interval that varies by ± 0.03 from the mean. With larger datasets, it would be possible to make a more conclusive comparison of whether adding age and gender features improves the system's ability to predict hypertension from fundus photographs and by how much.

To expand the proposed MMDL framework, future studies could integrate vessel segmentation methodologies, as demonstrated by Dai et al., who achieved a higher AUC on their CNN model by using a "segmented dataset" composed only of retinal blood vessels extracted with a pretrained U-Net based model (59). Furthermore, investigating multiclass classification problems to gauge deep learning models' ability to detect hypertension stage from fundus photographs and demographic features. Models that encompass hypertensive patients at various stages have the potential to uncover valuable insights regarding their capacity to detect hypertension in its early stages, as well as the areas of the image that were instrumental in generating such predictions.

Another limitation is the lack of diversity in the dataset, and including data from multiple races and underlying conditions could increase the generalizability of the studied models. To facilitate the integration of the proposed system into clinical environments, it is imperative to obtain additional data from across healthcare institutions to cover a wider range of patient demographics., serving to enhance the system's generalizability and overall performance.

**Conclusion**

We assessed four multimodal deep learning configurations and compared their performance to established benchmarks in the literature. We also trained a unimodal RETFound model on fundus photographs alone. By incorporating age and gender variables with the fundus photographs, we increased the F1 score to 0.77 with *HyMNet*, compared to 0.74 achieved by RETFound.

**Code Availability**

Both the source code and the model weights are openly available at:
https://github.com/MohammedSB/HyMNet.

**Data Availability**

The data can be requested through the following link:
https://kaimrc.ksau-hs.edu.sa/En/Pages/Ocular.aspx

**Declaration of generative AI and AI-assisted technologies in the writing process**

During the preparation of this work the author(s) used ChatGPT to enhance language and readability of this work. After using this tool/service, the author(s) reviewed and edited the content as needed and take(s) full responsibility for the content of the publication.

# HyMNet: a Multimodal Deep Learning System for Hypertension Classification using Fundus Photographs and Cardiometabolic Risk Factors

Supplementary Results

# A. Ablation Study

## A.1 Model Selection

We evaluated four models for the fundus path, two of which were CNN models and two were Vision Transformers. Specifically, we tested ResNet50 and DenseNet201 pre-trained on ImageNet1k, DINOv2 ViT-L/14 pre-trained on LVD-142M and RETFound pre-trained on 1.6 million retinal images from various sources. For each model, we tuned the learning rate for the newly initialized linear layer with the values {1e-2, 5e-2, 1e-3, 5e-3} and learning rates {1e-4, 5e-4, 1e-5, 5e-5, 1e-6, 5e-6} for the pre-trained backbone. In Supplementary Table 1, we present the AUC for linear-probing and fine-tuning for the best-performing checkpoint across the learning rates. We used 25 epochs training epochs for each run.

**Supplementary Table 1.** We report the results for hypertension classification for linear probing and fine-tuning two CNN and ViT models, pre-trained on natural datasets. Only the AUC for the best performing hyperparameter and checkpoint is shown.

| Model | Linear Probing | Fine-Tuning |
|---|---|---|
| ResNet50 | 0.650 | 0.690 |
| DenseNet201 | 0.643 | 0.677 |
| DINOv2 | 0.655 | 0.676 |
| RETFound | **0.685** | **0.695** |

## A.2 Effect of Image Size

In this section, we study the effect of image size on hypertension prediction ability. We evaluate the difference between processing fundus photographs at sizes 224x224 and 512x512 using RETFound on the test set. We observe a slight performance increase across F1, AUC, PR, and Accuracy scores, and a relatively larger increase for Recall. For this reason, we continued to use image size 512x512 for processing fundus photographs.

**Supplementary Table 2.** We evaluate the image size has on the ability to classify hypertension. Increasing the image size from 224x224 to 512x512 results in slightly better performance.

| Image Size | F1 | AUC | PR | Accuracy | Precision | Recall |
|---|---|---|---|---|---|---|
| 224 | 0.73 | 0.70 | 0.8 | 0.65 | **0.68** | 0.79 |
| 512 | **0.75** | **0.71** | **0.81** | **0.67** | **0.68** | **0.84** |

## A.3 Joint Fusion Feature Vector Size

We attempt to find the optimal sizes for the fundus and demographic paths in the joint fusion architecture. We scan {8, 32, 128} feature vector sizes for both paths, and report our results in Supplementary Figure 1. For the fundus path, we experimented with using the patch embeddings ("Patch") of the vision transformer, without an added trainable linear layer, as an output. Our experiments conclude that using a feature vector size of 8 for the fundus path and a size of 32 for the demographic best performs the best. We use RETFound for the fundus path.

| Demo \ Fundus | Patch | 8 | 32 | 128 |
|---|---|---|---|---|
| 8 | 0.658 | 0.647 | 0.643 | 0.659 |
| 32 | 0.648 | 0.67 | 0.65 | 0.664 |
| 128 | 0.603 | 0.664 | 0.657 | 0.649 |

**Supplamentry Figure 1. Feature embedding matrix.** We report the AUC for using different feature vector sizes for the fundus path "Fundus" and demographic path "Demo." As show in the figure, an embedding of size 8 for the fundus path and size 32 for the demographic path results in the optimal AUC.

# Supplementary Figure 2.

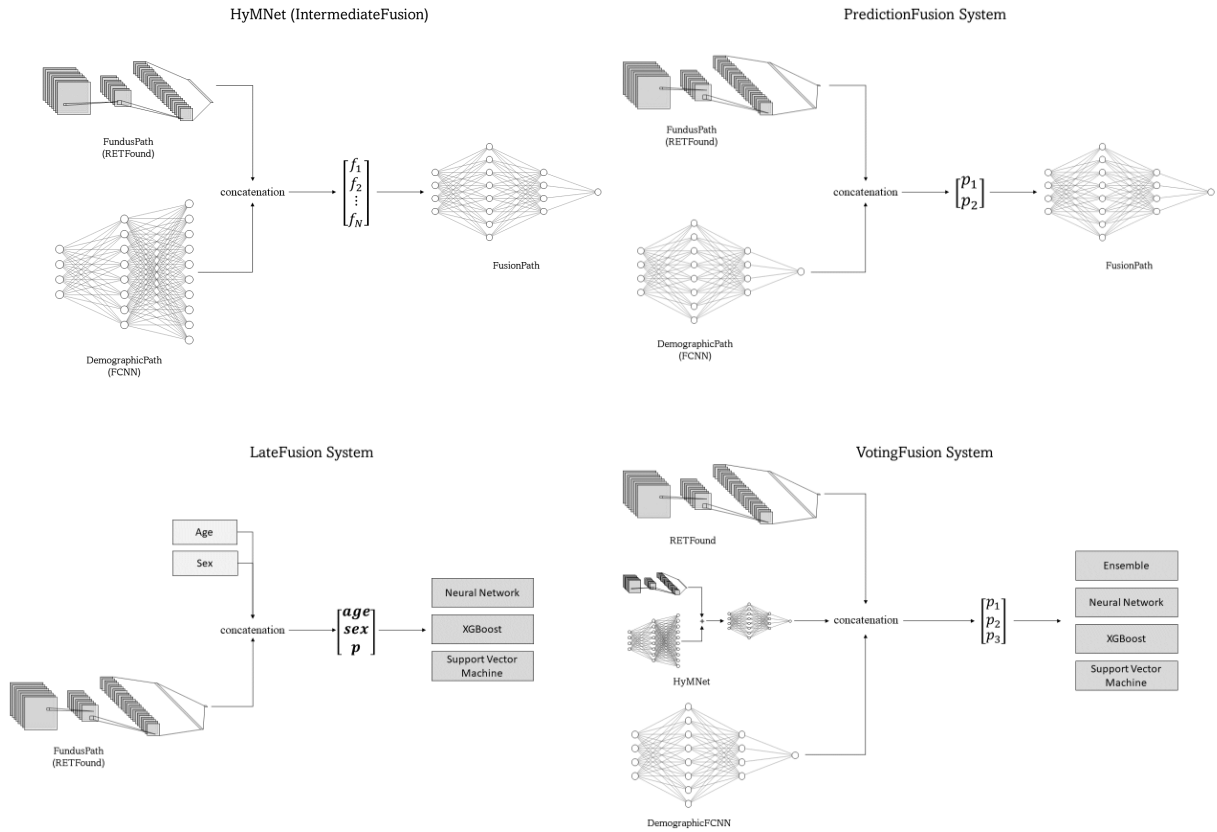

**Supplementary Figure 2. Multimodal systems diagram.** On the top left, the IntermediateFusion system is shown, where deep feature outputs from the FundusPath and the DemographicPath, represented as $f_1 - f_N$, are concatenated and fed into the FusionPath. The top right diagram shows the PredictionFusion system, where predictions logits, $p_1$ and $p_2$, are concatenated instead. The bottom two diagrams show the LateFusion and VotingFusion systems. In LateFusion, age and gender features are concatenated to the prediction logit of a fully trained FundusClassifierPath and passed into three different classifiers. Lastly, in VotingFusion, the prediction logit of a fully trained FeatureFusion system is concatenated with the prediction logits of a trained FundusClassifierPath and a trained DemographicClassifierPath.

# Supplementary Table 3.

**Supplementary Table 3. Multimodal and unimodal systems results.** The table presents the F1-score, AUC, PR, precision, recall, and specificity scores for all multimodal and unimodal systems used. CI represents the 95% confidence interval generated from the bootstrap technique mentioned in Section 2.5. A classification threshold of 0.5 was used for the F1-score, precision, recall, and specificity.

| Model | F1 | AUC | PR | Accuracy | Precision | Recall |
|---|---|---|---|---|---|---|
| Multimodal Systems | | | | | | |
| HyMNet (IntermediateFusion) | 0.771 [0.747, 0.796] | 0.705 [0.672, 0.738] | 0.743 [0.703, 0.784] | 0.690 [0.662, 0.719] | 0.683 [0.65, 0.716] | 0.887 [0.862, 0.912] |
| PredictionFusion | 0.758 [0.733, 0.784] | 0.672 [0.637, 0.707] | 0.713 [0.671, 0.755] | 0.679 [0.646, 0.713] | 0.676 [0.647, 0.706] | 0.860 [0.833, 0.888] |
| LateFusionXGB | 0.758 [0.733, 0.783] | **0.717 [0.684, 0.75]** | 0.749 [0.708, 0.791] | 0.685 [0.652, 0.719] | 0.68 [0.652, 0.708] | 0.848 [0.819, 0.877] |
| LateFusionSVM | 0.768 [0.744, 0.792] | 0.685 [0.652, 0.718] | 0.732 [0.692, 0.773] | 0.657 [0.625, 0.689] | 0.67 [0.641, 0.699] | 0.925 [0.905, 0.946] |
| LateFusionFCNN | 0.751 [0.726, 0.777] | 0.697 [0.664, 0.73] | 0.742 [0.703, 0.782] | 0.682 [0.649, 0.715] | 0.673 [0.644, 0.702] | 0.836 [0.807, 0.866] |
| VotingFusionXGB | 0.745 [0.718, 0.772] | 0.704 [0.671, 0.737] | 0.751 [0.712, 0.79] | 0.680 [0.646, 0.715] | 0.667 [0.639, 0.696] | 0.823 [0.792, 0.854] |
| VotingFusionSVM | 0.770 [0.747, 0.794] | 0.696 [0.663, 0.73] | 0.746 [0.707, 0.786] | 0.644 [0.613, 0.675] | 0.662 [0.634, 0.691] | **0.960 [0.945, 0.976]** |
| VotingFusionFCNN | 0.739 [0.713, 0.766] | 0.682 [0.648, 0.716] | 0.735 [0.696, 0.775] | 0.674 [0.64, 0.708] | 0.659 [0.63, 0.688] | 0.82 [0.789, 0.851] |
| VotingFusionEnsemble | **0.772 [0.748, 0.796]** | 0.712 [0.68, 0.744] | 0.748 [0.708, 0.788] | **0.692 [0.664, 0.72]** | **0.688 [0.655, 0.721]** | 0.881 [0.856, 0.906] |
| Unimodal Systems | | | | | | |
| RETFound | 0.745 [0.719, 0.772] | 0.690 [0.657, 0.724] | 0.740 [0.701, 0.78] | 0.682 [0.647, 0.717] | 0.668 [0.639, 0.698] | 0.821 [0.791, 0.852] |
| DemographicXGB | 0.756 [0.73, 0.782] | 0.697 [0.665, 0.73] | 0.736 [0.695, 0.778] | 0.684 [0.655, 0.713] | 0.696 [0.662, 0.73] | 0.827 [0.797, 0.858] |
| DemographicSVM | 0.765 [0.74, 0.79] | 0.706 [0.674, 0.738] | **0.752 [0.713, 0.791]** | 0.671 [0.642, 0.7] | 0.661 [0.629, 0.695] | 0.907 [0.884, 0.93] |
| DemographicFCNN | 0.752 [0.727, 0.778] | 0.694 [0.661, 0.727] | 0.742 [0.703, 0.782] | 0.661 [0.632, 0.69] | 0.662 [0.63, 0.695] | 0.871 [0.845, 0.898] |